\documentclass[10pt,a4paper]{iopart}
\makeatletter
\def\ps@myheadings{\let\@oddfoot\@empty\let\@evenfoot\@empty
    \def\@oddhead{\parbox{\textwidth}{\scriptsize This is an author-created, un-copyedited version of an article accepted for publication in \emph{Journal of Physics B: Atomic, Molecular and Optical Physics}. IOP Publishing Ltd is not responsible for any errors or omissions in this version of the manuscript or any version derived from it. The definitive publisher authenticated version is available online at {\ttfamily 10.1088/0953-4075/42/24/245302}}}%
    }
\makeatother
\usepackage{iopams}  
\usepackage[latin1]{inputenc}
\usepackage[T1]{fontenc}
\usepackage{pifont}
\usepackage{textcomp}
\usepackage{mathptmx}
\usepackage{graphicx}

\begin{document}
\title[Continuous loading of an optical dipole trap with magnetically guided ultra cold atoms]{A proposal for continuous loading of an optical dipole trap with magnetically guided ultra cold atoms}
\author{A Aghajani-Talesh, M Falkenau, A Griesmaier and T Pfau}
\address{Universität Stuttgart, 5.~Physikalisches Institut, Pfaffenwaldring 57\\ D-70550 Stuttgart, Germany}
\ead{a.aghajani@physik.uni-stuttgart.de, t.pfau@physik.uni-stuttgart.de}
\pacs{37.10.Gh, 37.20.+j}
\vspace{2pc}
\noindent{\it Keywords}: Cold atoms, Atom guides, Atom traps, Optical traps, Laser Cooling

\begin{abstract}
The capture of a moving atom by a non-dissipative trap, such as an optical dipole trap, requires the removal of the excessive kinetic energy of the atom. 
In this article we develop a mechanism to harvest ultra cold atoms from a guided atom beam into an optical dipole trap by removing their directed kinetic energy. 
We propose a continuous loading scheme where this is accomplished via deceleration by a magnetic potential barrier followed by optical pumping to the energetically lowest Zeeman sublevel. 
We theoretically investigate the application of this scheme to the transfer of ultra cold chromium atoms from a magnetically guided atom beam into a deep optical dipole trap. 
We discuss the realization of a suitable magnetic field configuration. 
Based on numerical simulations of the loading process we analyze the feasibility and efficiency of our loading scheme.
\end{abstract}

\section{Introduction}

The production of a Bose-Einstein condensate (BEC) is typically performed by a time sequence of cooling steps leading to an average yield of 10$^4$ to 10$^6$ atoms/sec for the best alkali experiments and $10^3$ atoms/sec for chromium BEC \cite{Griesmaier05}.
In order to overcome the inherent limitations of sequential BEC production processes, considerable effort has been invested into alternative concepts based on magnetic guides loaded with ultra-cold atoms \cite{Mandonnet00}. Magnetic guides allow to spatially separate and to perform simultaneously otherwise interfering cooling steps. They offer thus the prospect to prepare BECs continuously with considerably increased production rates. A spectacular application of this would be the realization of a truly continuously pumped atom laser \cite{SPREEUW95}, the matter wave analogue to an optical cw laser. Atom lasers could so far only be realized in pulsed and quasi-continuous mode \cite{Anderson98,Mewes97,Bloch99,Chikkatur02,Hagley99,Cennini03,Robins08}. A continuous atom laser would constitute a uniquely bright and coherent source of quantum matter with applications ranging from fundamental research on atomic physics to microscopy and lithography. 

Several methods have been demonstrated to load a magnetic guide from a magneto optical trap (MOT), resulting in loading rates of up to $7 \cdot 10^9$ atoms/sec \cite{Lahaye04}. For chromium, our group has recently reported continuous loading of $>10^9$ atoms/sec by operating a moving molasses MOT in the field of a magnetic guide \cite{Griesmaier09}. This is a good starting point for a continuously refilled reservoir of atoms. Efficient transfer of the atomic flux from a guide into an optical dipole trap (ODT) would open the possibility to reach degeneracy via evaporative cooling inside the ODT. It would thus constitute an important step towards high flux and even continuous BEC production. 

The capture of a moving atom by an ODT requires, due to the conservative character of the ODT potential, that the excessive kinetic energy of the atom is removed. Typical atom velocities in a guide are on the order of few m/s, the amount of kinetic energy that needs to be dissipated therefore can exceed typical trap depths of an ODT by more than one order of magnitude. In this article we propose and analyze a loading method which employs deceleration by an additional magnetic field inside the ODT region and subsequent optical pumping to the energetically lowest Zeeman-sublevel. 
This process bears similarities to a single Sisyphus-cooling cycle \cite{MetcalfBk99,Gauck98,Ruschhaupt04,Raizen05}. The feasibility and efficiency of this method is investigated by numerical simulations.

\section{Principle of the ODT loading mechanism}

\begin{figure}
\begin{center}
\includegraphics[height=4.5 cm]{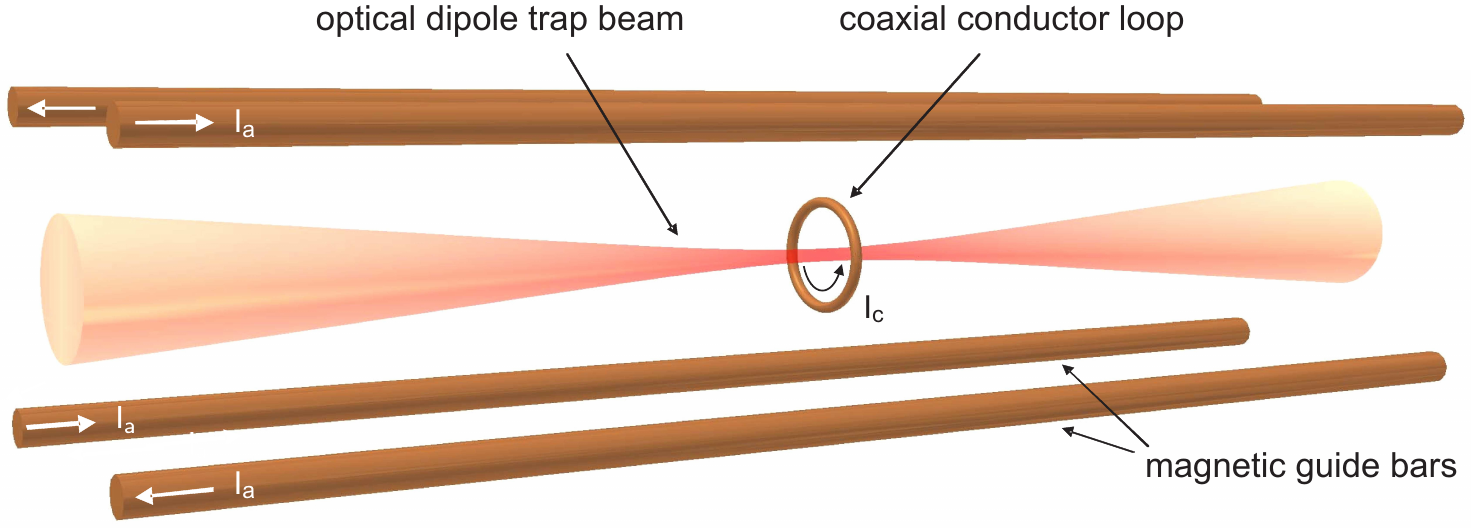}
\end{center}
\caption{Illustration of the trap geometry we propose for continuous loading of an ODT from a magnetic guide. The current $I_a$ that flows through the four bars in alternating directions generates a 2D magnetic quadrupole field which constitutes a magnetic guide for ultra-cold chromium atoms. The ODT is produced in the focus of an intense far red detuned laser beam  that is coaxially aligned with the axis of the atom guide. 
Atoms approaching the centre of the ODT are decelerated by the magnetic field of a coaxial conductor loop that carries the current $I_c$. }
\label{fig:Bars}
\end{figure}

The loading mechanism we propose is ideally applicable to chromium atoms, which due to their relatively high ground state magnetic moment of 6~$\mu_B$ are especially well suited for magnetic guiding. Recent experiments in our group have demonstrated the injection of ultra-cold $^{52}$Cr atoms out of a moving molasses MOT \cite{Cren02} into a horizontal magnetic guide \cite{GreinerA07}. A continuous flux of up to 6$\cdot$10$^9$ atoms/sec with tuneable velocities between 2 and 20 m/s and temperatures ranging from 1-2 mK could be achieved with this setup  \cite{Griesmaier09}. A goal of this project is the continuous transfer of the atomic flux into a deep ODT generated by an intense fibre laser.

In the following, we develop our theoretical considerations in close reference to our experimental setup. 
The magnetic guide consists of four parallel bars with rectangularly arranged centres (see figure~\ref{fig:Bars}). We choose the coordinate system such that the axis of the guide, its axis of symmetry, coincides with the z--axis. The four bars intersect the xy--plane at the points $(d/2,d/2,0)$, $(d/2,-d/2,0)$, $(-d/2,d/2,0)$ and $(-d/2,-d/2,0)$, respectively. Here $d$ corresponds to the distance of two neighbouring bars. A current $I_{\mathrm{a}}$ flows through the bars in alternating opposing directions. The resulting magnetic field $\mathbf{B}_{\mathrm{a}}$ inside the atom guide can be well approximated by a 2D quadrupol field:
\begin{equation}
\label{Equ:BGuide}
\mathbf{B}_{\mathrm{a}}(\mathbf{r}) = \frac{ 4  \mu_0 I_{\mathrm{a}} }{ \pi d^2 } (-x, y, 0)
\end{equation}
An atom interacts with the field via its magnetic moment. Its resulting potential energy $U_a$ depends only on the radial distance $\rho = \left(x^2+y^2\right)^{1/2}$ from the guide axis. With equation~\ref{Equ:BGuide} its potential energy it can be expressed as
\begin{equation}
\label{Equ:PotMag}
U_\mathrm{a}(\rho) =  \mu_\mathrm{B} g_J m_J \left| \nabla \left|\mathbf{B}_\mathrm{a}\right| \right| \rho,
\end{equation}
with $\mu_\mathrm{B}$ representing the Bohr magneton, $g_J$ the Landé g-factor and $m_J$ the magnetic quantum number of the atom, respectively. The term $\left| \nabla \left|\mathbf{B}_\mathrm{a}\right| \right|$ is a constant here and denoted as `magnetic field gradient'.  The guide potential described by equation~\ref{Equ:PotMag} leaves the atoms unconfined in the axial direction. It scales linear with $\rho$ and can, depending on the value of $m_J$, either be attractive, repulsive or equal zero. Thus, only atoms in specific Zeeman substates, so called `low field seeking states', are confined by the guide.

In order to explain the transfer of atoms from the magnetic guide into an ODT we assume that the ODT is 
generated by a far red detuned Gaussian laser beam with focal waist $w_0$ and Rayleigh range $z_\mathrm{R}$. We further assume that the optical axis of the beam coincides with the guide axis and that the focus is located at the origin of our coordinate system. For a total beam power $P_0$ the trap potential $U_\mathrm{d}$ of the ODT is then given by
\begin{equation}
U_\mathrm{d}(\mathit{\mathbf{r}})=
\frac{ P_0 \kappa h }{w_0^2} \left(1+\frac{z^2}{z_\mathrm{R}^2}\right)^{-1}
\exp{\left( -\frac{2 \rho^2 }{w_0^2} \left(1+\frac{z^2}{z_\mathrm{R}^2}\right)^{-1} \right)},
\label{Equ:ODT}
\end{equation} 
where $h$ is Planck's constant and the parameter $\kappa$ describes the coupling between the trap beam and the trapped atom \cite{Grimm98}. In the following, we regard $^{52}$Cr ground state atoms trapped by a laser beam with 300~W total power, 1070~nm wavelength and a focal waist of 30~\textmu{}m. For these settings equation~\ref{Equ:ODT} yields an optical trap depth of 3.6~mK.   

From equation~\ref{Equ:ODT} it can be seen that the extension of the ODT in the axial direction is essentially marked by $z_\mathrm{R}$ and in the radial direction by $w_0$. An atom can be regarded as trapped inside the ODT when the sum of its kinetic energy and its potential energy is lower than the ODT potential threshold. In order to load an atom into the ODT we therefore require a process that does not only remove a sufficient amount of its initial kinetic energy. In addition, the removal has to take place in close proximity to the centre of the ODT. 

Since only atoms that are in a low-field seeking Zeeman substate are transmitted inside the magnetic guide, an additional magnetic field can be used to generate a magnetic potential barrier inside the ODT that exerts a repulsive force on an atom approaching it from the guide. An atom that is decelerated by the barrier converts kinetic energy into potential energy up to the point where it either transcends the barrier or reaches a turning point of its trajectory. The position and orientation of the barrier with respect to the guide and the ODT have to be chosen such that this point of minimal axial kinetic energy lies as close as possible to the centre of the ODT. A fast optical pumping process can then be used to invert the orientation of the magnetic moment such that the decelerated atom is prepared with minimal kinetic energy in a high field seeking state. The formerly axially repulsive magnetic potential hence becomes attractive. The optical pumping process thus removes the potential energy that the atom has gained during the deceleration by the barrier. Depending on the amount of the remaining kinetic energy the atom would then be trapped in the combined potentials of the ODT and the magnetic field. 

\begin{figure}
\begin{center}
\begin{minipage}[b][5 cm][t]{2 ex} \small a) \vspace*{\fill} \end{minipage} 
\includegraphics[height=5 cm]{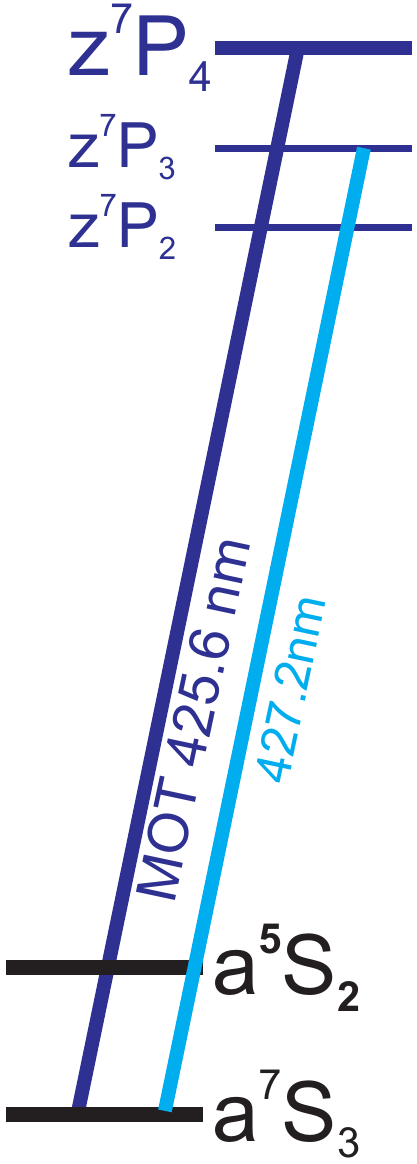}
\hspace{2 ex}
\begin{minipage}[b][5 cm][t]{3 ex} \small b) \vspace*{\fill} \end{minipage}
\includegraphics[height=5 cm]{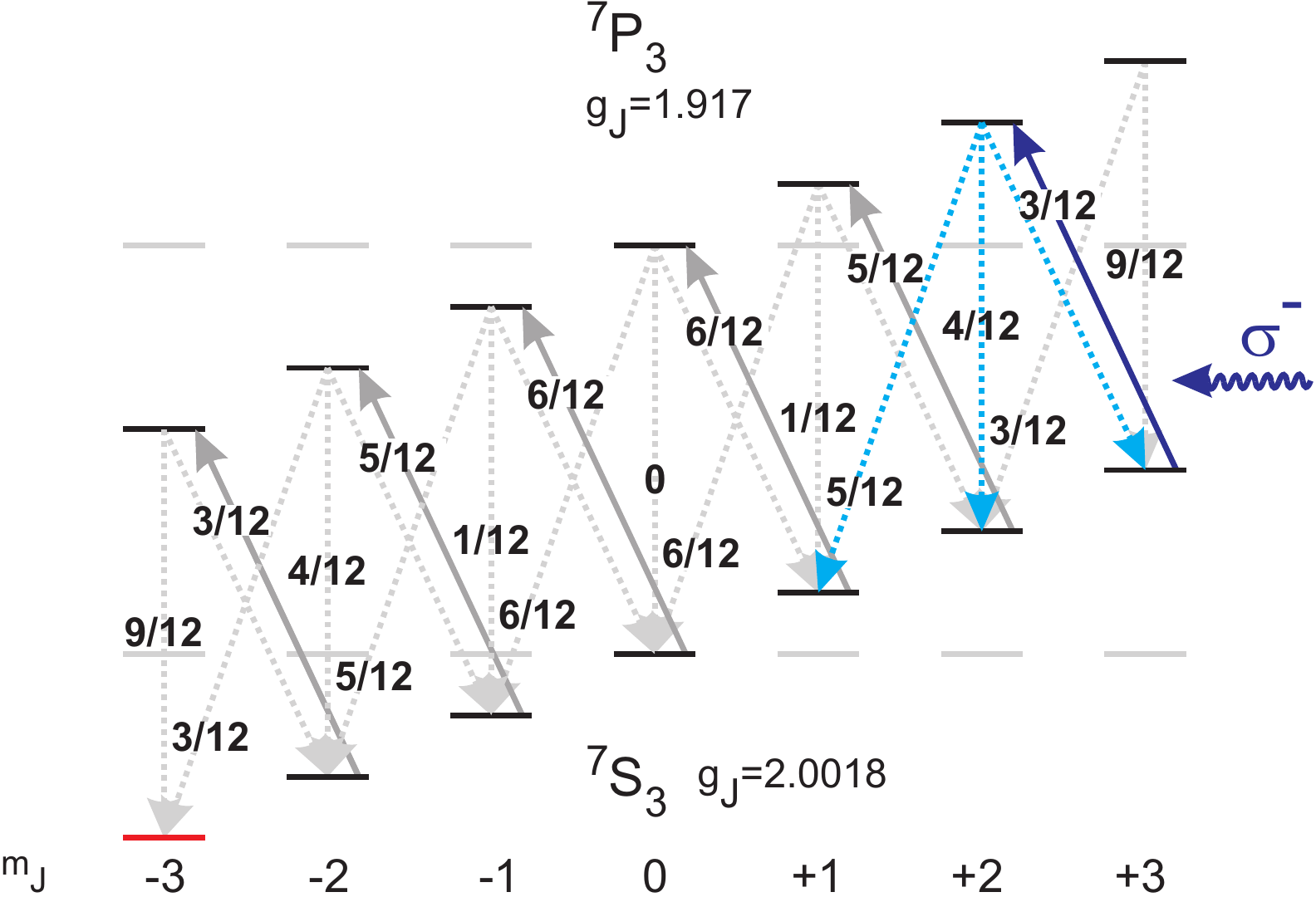}
\end{center}
\caption{a) $^{52}$Cr energy levels (not to scale) with optical transitions used for laser cooling and optical pumping. Atoms are guided in the $^7S_3$  ground state. b) Optical pumping to the $^7P_3$ state can be used to transfer atoms successively from $m_J=+3$ to the $m_J=-3$ Zeeman sublevel of the ground state.  
}
\label{fig:CrTerms}
\end{figure}

For our proposed loading scheme to be effective, it requires a suitable optical transition that allows the optical pumping to proceed both within a period much shorter than the average length of stay near the barrier peak and without causing excessive heating of the trapped atoms. For $^{52}$Cr, a transition that fulfils these requirements is presented by the $^7\mathrm{S}_3 \rightarrow ^7\mathrm{P}_3$ transition, which is depicted in figure~\ref{fig:CrTerms}~(a). The excited state has a lifetime of 33~ns. It decays with a branching ratio $> 10^3:1$ into the ground state. Figure~\ref{fig:CrTerms}~(b) illustrates subsequent optical pumping cycles driving an atom from the  $^7S_3$ $m_J=+3$ to the $m_J=-3$ level, the 
latter being a dark state. When $\sigma^{-}$ light is used, on average 6.2 photons are scattered during a single pumping cycle, causing negligible recoil induced heating \cite{GriesmaierPhD}.    

\begin{figure}
\begin{center}
\includegraphics[height = 7 cm]{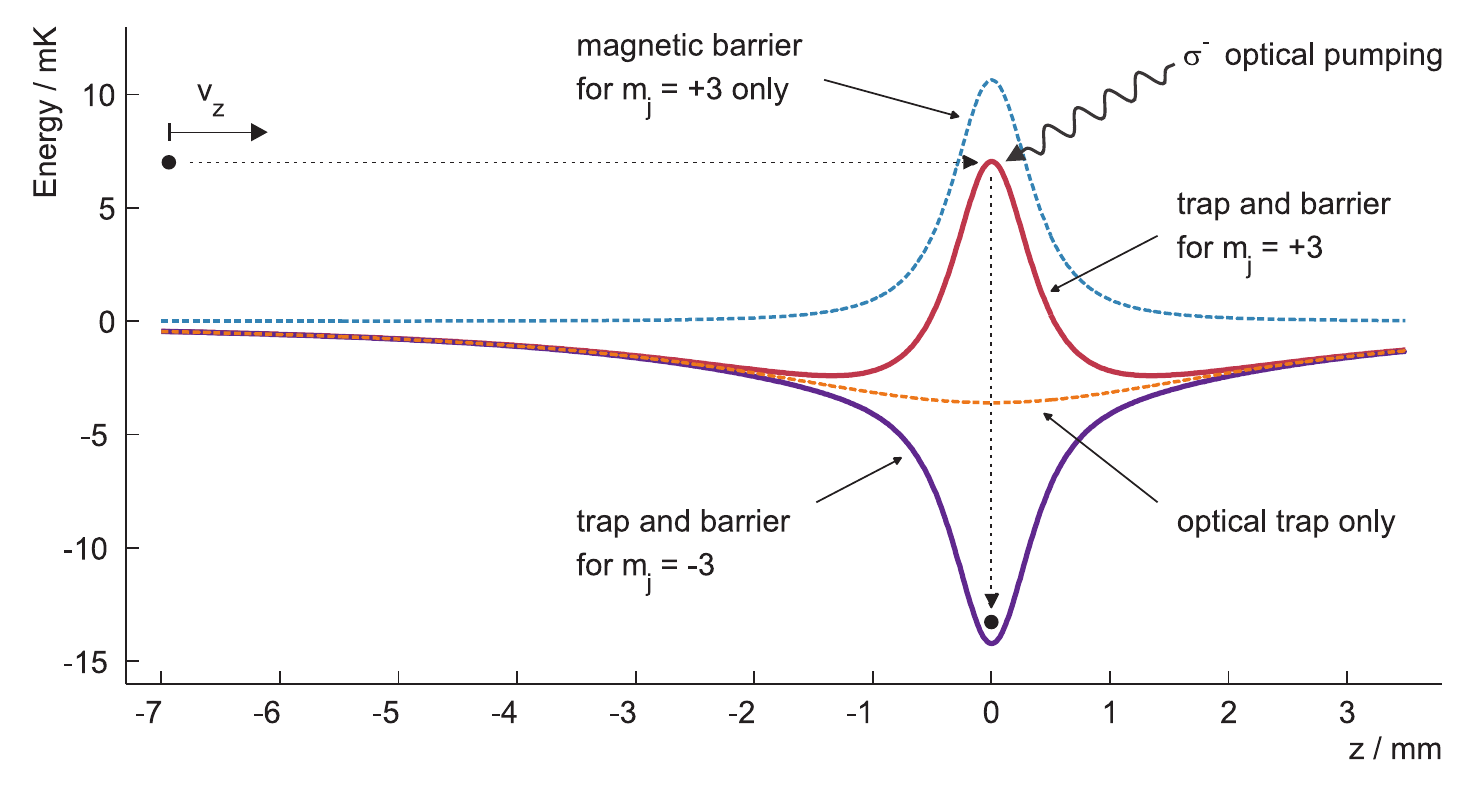}
\end{center}
\caption{Illustration of the proposed scheme for continuously loading the ODT from the atom guide. A Cr atom in the low-field seeking $m_J=+3$ ground state approaches the centre of the ODT and is decelerated by a magnetic potential barrier. The atom thereby converts kinetic into potential energy. At the centre of the ODT, where the atom has minimized its kinetic energy, it enters a beam of $\sigma^{-}$-polarized light, which pumps the atom to the high-field seeking $m_J=-3$ Zeeman sublevel. The pumping removes the gained potential energy and leaves the atom close to the bottom of a trap potential, which is generated by the ODT and the now attractive magnetic potential. Depending on the remaining kinetic energy, the atom remains trapped.  
}
\label{fig:Scheme}
\end{figure}

Figure~\ref{fig:Scheme} illustrates the loading scheme for $^{52}$Cr. An atom in the $m_J=+3$ sublevel of the $^7S_3$ ground state moves towards the ODT and is subsequently stopped at the centre of the ODT by the magnetic barrier superimposed with the ODT. At the point of return it enters a $\sigma^{-}$-polarized pump laser beam that drives the atom quasi-instantaneously to the $m_J=-3 $  $\ ^7S_3$ level. As a result, the atom is trapped by the ODT and the inverted magnetic potential.

\section{Magnetic field configuration and trap potentials}

\begin{figure}
	\begin{center}
		\includegraphics[ height=7 cm]{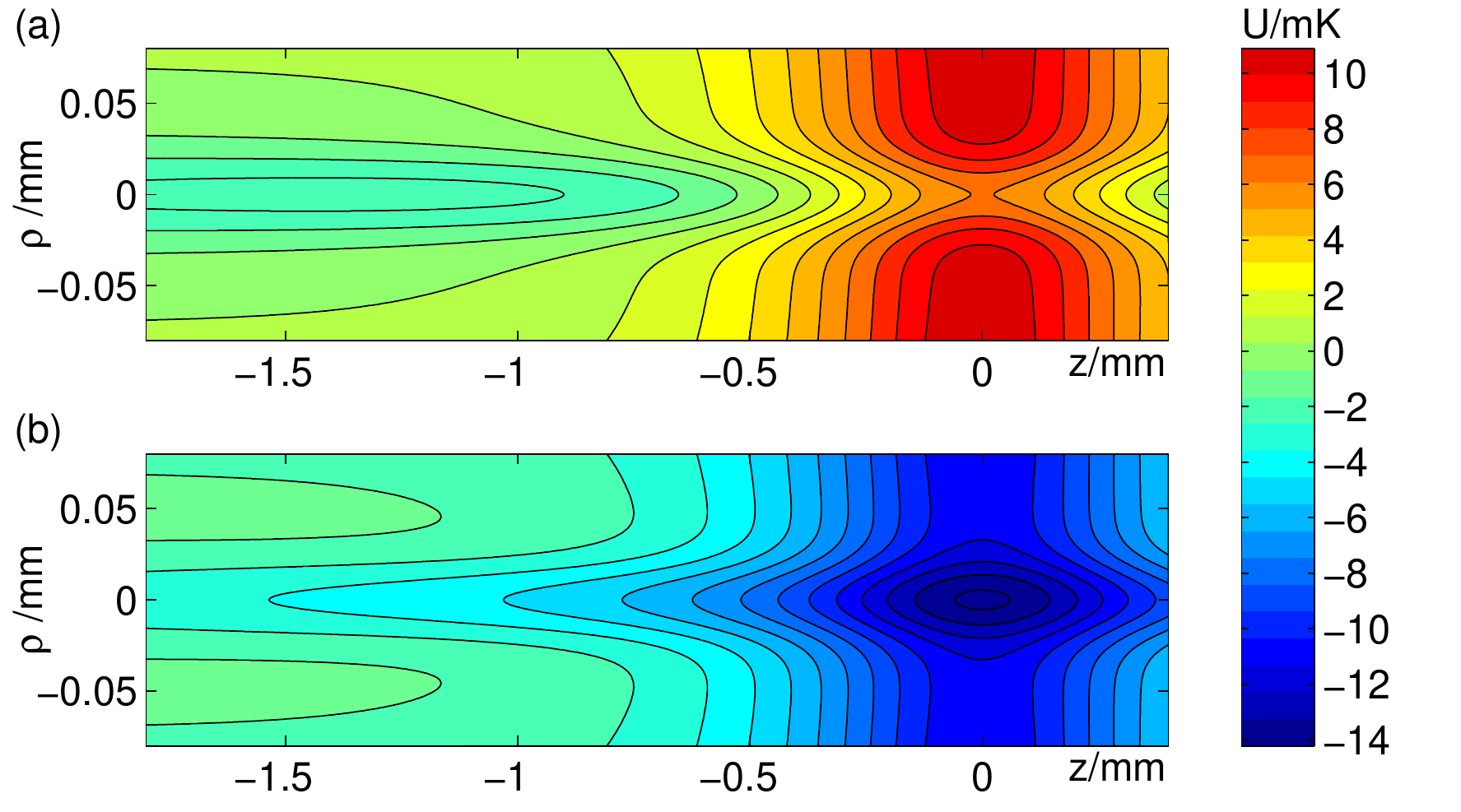} 
	 \end{center}
\caption{
Colour coded contour plots of the combined potentials of the atom guide, the optical dipole trap and the deceleration loop. Values are calculated for $R=0.5$~mm and $v_\mathrm{b}=1.5$~$\mathrm{ms}^{-1}$.
a) Potential for the low field seeking $m_J=+3$ state.  The potential increases along the z-axis toward the origin. Atoms moving towards the origin are decelerated. The combined potentials of the guide and the ODT provide radial confinement. b) Potential for the high-field seeking $m_J=-3$ state,  reached via after optical pumping. Atoms can be trapped at the origin in a potential well formed by the ODT and the decelerating magnetic field. 
}
\label{fig:Barrier}
\end{figure}

During the preceding description of the loading scheme we have implicitly assumed the existence of a suitably shaped magnetic potential barrier. For a further analysis of the loading scheme the magnetic field configuration has to be specified in more detail. It has to be considered here, that the field $\mathbf{B}_\mathrm{m}$ that determines the shape of the potential barrier is in general a vector sum of the guide field $\mathbf{B}_\mathrm{a}$ and other additionally applied fields. 
An important requirement for $\mathbf{B}_\mathrm{m}$ is that its orientation has to be uniform inside the region where the optical pumping takes place in order to allow the preparation of purely $\sigma^{-}$-polarized pump light. 
Moreover, while it is desirable to have a field whose strength increases steeply towards the ODT centre in axial direction, the curvature in radial direction has to be sufficiently small in order to avoid a defocusing effect on the atomic beam in either the low-field or the high-field seeking state.  

A suitable magnetic field can be obtained by adding a single circular current loop positioned concentrically with respect to the ODT and the magnetic guide, as shown in figure~\ref{fig:Bars}. 
The magnetic field $\mathbf{B}_\mathrm{c}$ produced by the current loop near the centre of the loop, which coincides with the ODT centre, is in line with the guide axis. Starting from the centre it increases slowly in radial direction and decays on the length scale of the coil diameter in the axial direction. A general analytical expression for $\mathbf{B}_\mathrm{c}(\mathbf{r})$ can be given in terms of elliptical integrals \cite{LehnerBk08}. 
Assuming the loop current $I_\mathrm{c}$ to be oriented positively with respect to the z-axis, $\mathbf{B}_\mathrm{c}$ is along the z-axis, for a loop radius $R$ given by   
\begin{equation}
\label{Equ:Biot}
\mathbf{B}_\mathrm{c}\left(z\right)= \frac{I_\mathrm{c} \mu_\mathrm{0}}{2} \ \frac{R^2 }{\left( R^2 + z^2 \right)^{3/2}} \ \ \mathbf{e}_z \ .
\end{equation}  
Having specified the magnetic field configuration we can express the potential $U_\mathrm{0}$, in which the atoms move during the loading, in terms of the fields of $\mathbf{B}_\mathrm{a}$ and $\mathbf{B}_\mathrm{c}$ and the ODT potential $U_\mathrm{d}$ as
\begin{equation}
U_\mathrm{0} = \mu_\mathrm{B} g_J m_J \left| \mathbf{B}_\mathrm{a} + \mathbf{B}_\mathrm{c} \right| + U_\mathrm{d} \ .
\label{Equ:potential}
\end{equation}    
The loop current has to be adjusted to the beam velocity $v_\mathrm{b}$ such that the total height of the potential barrier at the origin cancels the axial kinetic energy component of the arriving atoms. From equation~\ref{Equ:ODT} and \ref{Equ:Biot} follows thus that $I_c$ has to obey
\begin{equation}
\frac{m v_\mathrm{b}^2}{2} \ = \  \mu_\mathrm{B} g_J m_J \ \frac{I_\mathrm{c} \mu_\mathrm{0}}{2 R} \ + \ \frac{ P_\mathrm{0} \  \kappa h }{w_\mathrm{0}^2} 
\label{equ:barrierheight}
\end{equation} 

The equations above relate the geometry of the ODT and of the atom guide with the decelerating current loop. They can be used to find optimized values for the interdependent parameters $R$ and $I_c$. In a real experiment, however, there might be additional technical constraints that have to be considered. It might be  difficult for example to work with extremely small coils and rather large current in an ultra high vacuum (UHV) environment. Moreover, it might be necessary to use the same coils with a range of beam velocities and an accordingly wide range of loop currents. With the objective of providing useful guidance for the design of an experiment, we assume throughout this article a fixed loop radius of 0.5~mm and a magnetic field gradient of $350~\mathrm{G}\mathrm{cm}^{-1}$. 
From equation~{equ:barrierheight}, it then follows that the coil current $I_\mathrm{c}$ depends only on the beam velocity $v_\mathrm{b}$. 
In the following, we thus regard the magnetic barrier field and the resulting potential configuration as functions of $v_\mathrm{b}$ alone. For assumed beam velocities ranging from 1~ms$^{-1}$ to 5~ms$^{-1}$, the loop has to carry currents between $1.3$~A and $16.4$~A. With technologies adapted from magnetic micro traps it should be feasible to operate a coil subject to these requirements \cite{Fortagh07}.

Colour coded contour plots of total potential $U_\mathrm{d}$ in the $xz$-plane are shown in figure~\ref{fig:Barrier}. An atom in the low-field seeking $m_J=+3$ state experiences a potential barrier as shown figure~\ref{fig:Barrier}~(a). In contrast, an atom in the high-field seeking $m_J=-3$ state experiences a trap-shaped potential. In both cases the radial confinement is solely provided by the ODT. The depicted potentials correspond to a barrier height adjusted for a beam velocity of only 1.5~ms$^{-1}$. At higher velocities the magnetic barrier height becomes so large that the contribution from the ODT would be hardly recognizable.

\section{Simulation of the loading process} 

\begin{figure}
\begin{center}
\begin{minipage}[b][5.5 cm][t]{2 ex} \small a) \vspace*{\fill} \end{minipage} 
\includegraphics[height=5.5 cm]{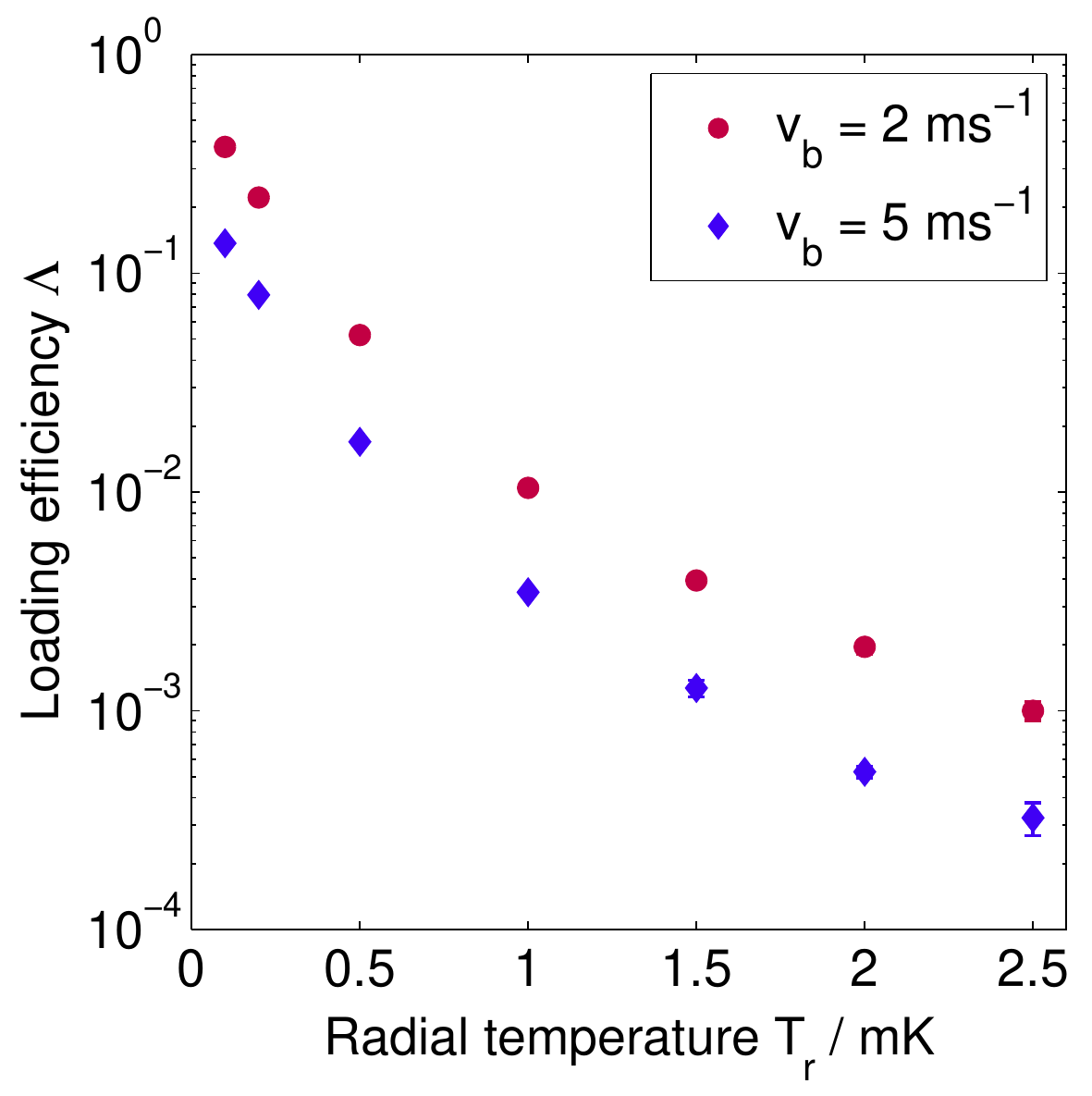}
\hspace{2 ex}
\begin{minipage}[b][5.5 cm][t]{3 ex} \small b) \vspace*{\fill} \end{minipage}
\includegraphics[height=5.5 cm]{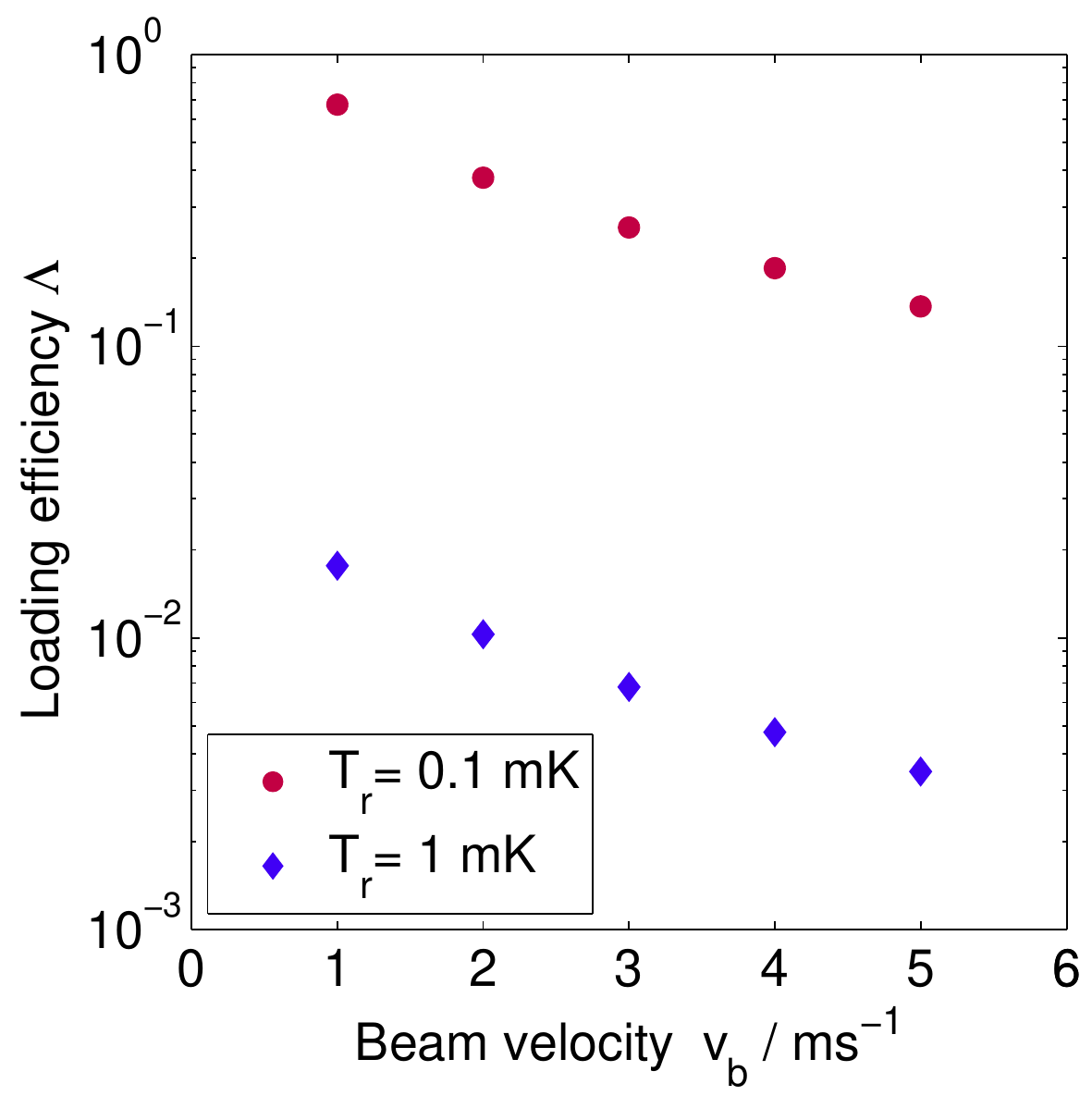}
\end{center}
\caption{ a)  Calculated loading efficiency $\Lambda$ as function of the initial radial temperature $T_\mathrm{r}$ of the atoms inside the guide. The velocities in the axial direction are assumed to be distributed according to a Maxwell-Boltzmann distribution with temperature $T_\mathrm{z}=1$~mK offset by a beam velocity $v_\mathrm{b}$ of 2~ms$^{-1}$ and 5~ms$^{-1}$, respectively. It becomes evident that a further reduction of $T_\mathrm{r}$ from 1~mK (currently achievable) to 125~\textmu K (Doppler-temperature of Cr) would offer the prospect of a substantial increase in the loading efficiency. b) Loading efficiency as a function of the beam velocity for an axial temperature of 1~mK and values of the radial temperatures of 0.1~mK and 1 mK, respectively. A reduction of the beam velocity within range of 1~ms$^{-1}$ to 5~ms$^{-1}$ would only lead to a moderate enhancement of the loading efficiency.     
}
\label{fig:SimRes1}
\end{figure}
In order to assess the feasibility and the efficiency of our proposed loading scheme we have numerically simulated the transfer of atoms from the guided atom beam into the ODT.  For a given configuration of the atom beam and the trapping potential we are interested in the overall loading efficiency $\Lambda$, which we define as ratio between the fraction $\Phi_\mathrm{L}$ of the atomic flux that remains trapped in the ODT after optical pumping and the total incoming flux $\Phi_\mathrm{0}$. 
From experimental studies it is known that the radial and axial distributions of atomic velocities and positions inside the chromium atom beam resemble thermal distributions. They can be well described by specifying the beam velocity $v_\mathrm{b}$, the total flux $\Phi_0$, and the respective radial and axial beam temperatures $T_\mathrm{r}$ and $T_\mathrm{z}$ \cite{Griesmaier09}. 
As outlined in the preceding section, we restrict our studies to a potential configuration that depends only on $v_\mathrm{b}$. Moreover, we make the important assumption that the densities in the atom guide and the ODT remain low enough such that inter-atomic collisions can be safely neglected during the loading process. It follows that, under these premises, $\Lambda$ is density independent and can be regarded as a function of $v_\mathrm{b}$, $T_\mathrm{r}$ and $T_\mathrm{z}$ alone.  

Our simulations of the transfer of atoms from the atom guide into the ODT are based on the computation of a large number of individual atom trajectories and their subsequent analysis. For the computation of the trajectories we have used the expression for the total potential energy stated in equation~\ref{Equ:potential} to obtain analytical expressions for the corresponding equations of motion. In the absence of collisions, we can for given initial conditions obtain the full 3D trajectories by numerically integrating the equations of motion on a personal computer using standard commercial mathematical software. 

We then use the resulting trajectory to estimate a likely position and velocity of the atom when it undergoes the optical pumping \cite{Power97}. For this purpose we require that the duration of the optical pumping process is negligibly short and that the pump beam is restricted to a narrow region near the top of the magnetic potential barrier. We then assume that as soon as the atom either has lost all its axial velocity ($v_\mathrm{b}=0$) or as soon as it reaches the maximum of the potential barrier ($z=0$) it is pumped instantaneously into the high-field seeking state, where it is suddenly subject to an attractive potential. We regard the atom as successfully transferred into the ODT if the optical pumping process takes place inside the region of the final trap potential well and if the remaining total energy of the atom is lower than the trap potential threshold.

For the computation of the trajectories the atoms are all initialized at $z = -0.05$~m, which is chosen to be sufficiently far away from the ODT-centre, such that the initial potential energy distribution is only determined by the guide potential. For the initial radial distribution $n\left( \rho \right)$ follows from equation~\ref{Equ:PotMag}
\begin{equation}
n\left( \rho \right) = \rho \beta^{-2} \exp\left[- \frac{ \rho}{\beta} \right],
\label{Equ:rdist}
\end{equation}
where 
\begin{equation}
\beta=\frac{k_\mathrm{B} T_\mathrm{r}}{ \mu_\mathrm{B} g_J m_J \left| \nabla \left|\mathbf{B}_\mathrm{a}\right| \right|}.
\label{Equ:beta}
\end{equation}
The initial radial velocities $v_x$ and $v_y$ are distributed according to  Maxwell-Boltzmann distributions with temperature $T_\mathrm{r}$. The distribution $\nu(v_\mathrm{z})$ of the longitudinal velocities is described by a Maxwell-Boltzmann distribution with temperature $T_\mathrm{z}$ centred around the beam velocity $v_\mathrm{b}$, yielding 
\begin{equation}
\nu(v_\mathrm{z}) = \sqrt{\frac{m}{2 \pi k_\mathrm{B} T_\mathrm{z}}} \ \exp\left[ - \frac{ m \left(v_\mathrm{z} - v_\mathrm{b}\right)^2 }{2 k_\mathrm{B} T_\mathrm{z}} \right]. 
\label{Equ:vz}
\end{equation}
In order to determine $\Lambda$ we have at each instance calculated a large number of trajectories ($> 5 \cdot 10^4$) with initial conditions randomly sampled according to the respective values of $T_\mathrm{r}$, $T_\mathrm{z}$ and $v_\mathrm{b}$. The number of successful transfers divided by the total number of trajectories then yields the corresponding value of $\Lambda$. 

The results of our simulations are presented in figure~\ref{fig:SimRes1}. 
We have concentrated our studies on the dependence of $\Lambda$ on the radial beam temperature (subfigure~\ref{fig:SimRes1}~(a) and on the beam velocity (subfigure~\ref{fig:SimRes1}~(b), respectively. With the objective to investigate the application of our loading scheme to a real experiment and to derive strategies for an optimization of the loading rate, we have in both cases regarded parameter ranges that we deem to be experimentally accessible with present atom beam preparation methods \cite{Griesmaier09,GreinerA07}. 
As $T_\mathrm{z}$ is typically roughly constant, we have used a, conservatively estimated, fixed value of $T_\mathrm{z}=1$~mK for the simulations presented in this article.

The dependence of $\Lambda$ on $T_\mathrm{r}$ is shown in figure~\ref{fig:SimRes1}~(a). Two data sets are displayed, corresponding to two different beam velocities: $v_\mathrm{b}=2~\mathrm{m s}^{-1}$ (red circles) and $v_\mathrm{b}=~5\ \mathrm{m s}^{-1}$ (blue diamonds). The value of $2~\mathrm{m s}^{-1}$ represents the minimal velocity at which the beam can be effectively operated, while at approximately $5~\mathrm{m s}^{-1}$ the beam yields maximum atom flux \cite{Griesmaier09}. Both data sets exhibit a steep increase of $\Lambda$ with decreasing $T_\mathrm{r}$. We attribute this to the energy initially stored in the transverse degrees of freedom, which cannot be dissipated with our loading scheme, and which contributes on average with $3~k_\mathrm{B} T_\mathrm{r}$ to the total energy remaining after optical pumping. 
The course of $\Lambda$ as a function of $T_\mathrm{r}$ indicates that additional measures to decrease $T_\mathrm{r}$, even if they are associated with a moderate reduction of $\Phi_\mathrm{0}$, might be helpful in order to maximize the loading rate. At $T_\mathrm{r}=1$~mK, which represents the presently lowest values for the radial beam temperature \cite{Griesmaier09}, the graph yields $\Lambda=$ 0.35~\% for $v_\mathrm{b} = 5 \mathrm{m s}^{-1}$. Doppler cooling in the radial direction might, for instance, be used to reduce the radial temperature further to 0.125~mK, the Doppler temperature of chromium. 
In this case the loading efficiency would be expected to increase from 0.35~\% to over 12~\%. 

From figure~\ref{fig:SimRes1}~(a) we can learn that for $v_\mathrm{b}=5~\mathrm{m s}^{-1}$ the loading efficiency is about three times smaller than for $v_\mathrm{b}=2~\mathrm{m s}^{-1}$. This can be explained by the different respective height of the magnetic potential barrier that is needed in order to decelerate the atoms. An increased barrier height results in an increased radial curvature und thus lowers the effective depth of the trapping potential. For this reason, our loading scheme can not be applied to arbitrarily high beam velocities. In figure~\ref{fig:SimRes1}~(b) the dependence of $\Lambda$ on $v_\mathrm{b}$ is shown in more detail for two different radial temperatures, $T_\mathrm{r}=1~\mathrm{mK}$ and $T_\mathrm{r}=0.1~\mathrm{mK}$. The results from the simulations have to be contrasted with the experimentally observed dependence of the total flux $\Phi_\mathrm{0}$ on $v_\mathrm{b}$, which exhibits a pronounced maximum between  $5$~ms$^{-1}$ and $6$~ms$^{-1}$ and a steep decrease towards slower beam velocities \cite{Griesmaier09}. Regarding the optimization of the loading rate, it appears to be evident that the decrease of $\Phi_\mathrm{0}$ towards smaller beam velocities could possibly not be compensated by a corresponding increase of $\Lambda$. Thus it might not be advisable to reduce the beam velocity below $5$~ms$^{-1}$ at the expense of a greatly reduced total flux. 

Based on the values of $\Lambda$ and the reported values of $\Phi_\mathrm{0}$ \cite{Griesmaier09} we estimate for a beam velocity of 
$5$~ms$^{-1}$ and a radial temperature of 1~mK a loading rate of about $4 \cdot 10^6$ atoms/sec. We estimate that under these conditions it would take less than 1~s to reach a steady state with more than $10^6$ atoms in the ODT.
If at this point the loading process would be interrupted, this would already provide excellent starting conditions for the application of demagnetization cooling \cite{Fattori06}. This technique provides highly efficient cooling without loss of atoms and has shown to work best in dense and hot clouds that provide high collision rates and require easily controllable magnetic fields.

\section{Conclusion}
In this article we have outlined a scheme for the transfer of ultra cold atoms from a magnetically guided atom beam into an optical dipole trap. The scheme compromises deceleration by a magnetic barrier superimposed with the ODT, followed by optical pumping to the energetically lowest Zeeman-sublevel. We have provided an elaborate discussion of the application of this scheme to a beam of ultra cold chromium atoms. 
The preparation of suitable potential field configurations for the deceleration and the trapping of the atoms have been treated. Using numerical simulations of the loading process have investigated the dependence
of the loading efficiency on the initial radial beam temperature and on the beam velocity. Our simulations suggest that, based on recently reported experimental data \cite{Griesmaier09}, loading efficiencies of about 0.35~\%, resulting in maximum loading rates of more than $10^6$ atoms/sec, are feasible.  

\ack
The authors of this article acknowledge financial support by the Landesstiftung Baden-Württemberg under contract No.~0904Atom08. AA-T acknowledges financial support by the Studienstiftung des Deutschen Volkes.\\

\bibliographystyle{unsrt}
\bibliography{PI5_AAT_JPhysB}

\end{document}